\documentclass[11pt,twoside]{article}
\usepackage{newpasp}

\index{Blitz, L.}
\begin{document}

\title{Local Group HVCs: Status of the Evidence}
\author{Leo Blitz}
\affil{Radio Astronomy Laboratory, University of California, Berkeley, CA 94720}
%\author{replace this with the 2nd authors name}
%\affil{replace this with the 2nd authors address}
%\author{replace this with the 3rd authors name}
%\affil{replace this with the 3rd authors address}

% A concise abstract is recommended.  Enter the text of the abstract in
% between the \begin{abstract} and \end{abstract} commands.  Do NOT
% include the word ``Abstract'' in your text; it is inserted
% automatically. Do NOT  make a paragraph break between \begin{abstract} 
% and the first line of the text of the abstract!  Abstracts are required 
% for all papers.

\begin{abstract}

	The evidence for locating the High Velocity Clouds in the Local
Group is summarized and evaluated.  Recent measurements of the H$\alpha$
surface brightness and metallicity of a number of HVCs appear to
be fatal to the Galactic fountain as a significant contributor to
the HVC phenomenon, but not to the existence of the fountain itself.
Observations of extragalactic analogues to HVCs remain the {\it sine
qua non} for deciding whether the Local Group hypothesis is viable,
but the constraints based on existing surveys appear to be rather weak.
MgII quasar absorption lines restrict how
many HVC analogues exist at intermediate redshift, depending on where these
lines originate.  It is concluded
that the evidence remains ambiguous, none of the main hypotheses
is fully consistent with all of the data, and the Local Group hypothesis 
remains a viable explanation for the HVC phenomenon.

\end{abstract}

% Include keywords if you wish. The keywords.apj file, found on aas.org 
% in the pubs/aastex-misc directory, contains a list of keywords used 
% with the ApJ and Letters.  


% That's it for the front matter.  On to the main body of the paper.

\section{Introduction}

	Several years ago, the old hypothesis that the High Velocity
Clouds (HVCs) are members of the Local Group was revived by Blitz et
al. (1999) in a modern cosmological context.  The revival of this idea
generated some interest in the community because the most formidable
objections to it were obviated by the introduction of dark matter,
and because the HVCs then played an important role in galaxy formation
and evolution.  If the idea were right, the HVCs would be imbued with
cosmological significance, and could be studied in some detail because
they are quite close at hand.  Most important, the Blitz et al. (1999)
study made a number of specific predictions that could be tested with
observations that are relatively straightforward, and it could be learned,
in principle, whether the idea is right or wrong in short order.

	Some of these tests were carried out, and it turned out
that the results, rather than clarifying the issue, added complications,
and the nature of the HVCs remains murky.
Some of these will be reviewed here to give a flavor of where things
stand as of this writing.  Space limitations preclude reviewing all of
the relevant observational material, but the two other most commonly
discussed hypotheses in the recent literature will also be addressed:
the Galactic fountain (Shapiro and Field 1976), and stripping from the
Magellanic clouds and other dwarf galaxies in the Local Group.

	It is worth noting that some authors point out that the
HVCs are likely to be a composite phenomenon, with more than one origin,
and the discussion of the HVCs ought to take this diversity into
account.  This view begs the issue on two counts.  First, whether or not
the HVCs have a single origin, there is likely to be one {\it dominant}
origin for the HVCs responsible for either most of the mass or most
of the individual catalogued clouds, and the application of Occam's
razor demands that we know what that dominant origin is.  Perhaps more
important, if the clouds are of Local Group origin, then they almost
surely have cosmological significance, and they must have counterparts
throughout the Universe.  Thus the real question is whether the HVCs play
a significant role (past or present) in galaxy formation and evolution,
or whether they are simply curiosities related to the Milky Way alone.

\section{The Modern Local Group Hypothesis}

	Of the various pieces of evidence for and against the numerous
hypotheses proposed to explain the dominant origin of the HVCs, most
are rather weak and ambiguous (e.g. Wakker \& van Woerden 1997), and
each hypothesis is probably more notable for its weaknesses rather than
for its strengths.  The modern Local Group hypothesis, however, has the
advantage of deriving from a simple dynamical argument that explains
the most fundamental aspects of the available data. 
Using the relatively
complete HVC catalogue of Wakker and van Woerden (1991), Blitz et
al. (1999) showed that a single compact hypothesis could explain both
the spatial and velocity distribution {\it simultaneously} of HVCs over
the entire sky.  If the Local Group hypothesis turns out to be incorrect,
the competing ideas will have to reproduce both the distribution on the sky
and the kinematics of the clouds in a straightforward manner, something
none of them has been able to do so far.

	The model identifies the HVCs with the earliest structures to
form in the Universe, and they are thus necessarily dark matter dominated.
Using only the gravity of the Milky Way and M31 (with minor
modifications), the model reproduces the observed spatial concentrations
on the sky, the shape of the envelope of observed velocities, the
amplitude of the distribution and the preponderance of higher absolute
velocities in the northern hemisphere.  Because the model is so simple,
it was possible to make several predictions related to the H$\alpha$
surface brightness of the clouds, metallicities, and the detectability of
extragalactic analogues, and to contrast these with the Galactic fountain
model, the model favored by most writers on the subject during the 1990s.

\section{Tests of the Local Group Hypothesis}

\subsection{The H$\alpha$ Test}

	In the late 1990s, H$\alpha$ had been detected toward the
largest of the HVCs, and toward the Magellanic Stream (MS -- Weiner \&
Williams 1996; Tufte, Reynolds \& Haffner 1998; Bland-Hawthorn et
al. 1998).  The largest HVCs are close to the Milky Way in all
models, and have a mean distance of about 10 -- 20 kpc in the LG
hypothesis, consistent with direct distance determinations along two
lines of sight (Danly, Albert \& Kuntz 1993; van Woerden et al. 1998).
There was some debate initially about whether the source of excitation
was photoionization from escaping UV radiation from the Galactic plane,
or shock heating by passage of the clouds through a tenuous Galactic
halo. In the Local Group hypothesis, most of the HVCs are at distances
larger than 50 kpc and should exhibit H$\alpha$ surface brightnesses
smaller than the weakest of the H$\alpha$ detections regardless of the
excitation mechanism.  Local Group HVCs will have either lower incident
ionizing radiation than the detections for the MS, or they will impinge
lower density halo gas, but in either case, the H$\alpha$ emitted from
the HVCs should have lower surface brightness.

	Several sets of observations were carried out, but the
most extensive to be published to date are those of Weiner, Vogel and
Williams (2001, this volume) who observed HVCs in the southern hemisphere.
It had been expected that most of the HVC detections would either be
considerably fainter than the measurements of the MS, if the HVCs are
Local Group objects, or much brighter, if they are part of a Galactic
fountain.  By calibrating the measured surface brightness to a cloud
or clouds of known distance, absolute distances to the clouds can 
be determined by tying the calibration to either a model of the
ionizing radiation escaping from the Galaxy or to estimates of the halo
density if the HVCs are shock heated.

	Judging from Figure 2 of Weiner et al., the situation is much
more complex.  It had been hoped that the MS, with its known distance,
would serve as a calibration for the H$\alpha$ observations, but it turned
out that observed fluxes toward the MS vary by two orders of magnitude,
making the stream all but useless for calibration purposes.  Second, two HVCs,
complexes A and M, have measured distances smaller than the MS clouds,
yet their H$\alpha$ surface brightness is {\it fainter} than many of
the MS lines of sight.  Using a conservative photoionization model and
using clouds A and M as calibrators, Weiner et al. conclude that the
fainter HVCs they observed have distances inconsistent with an origin in
a Galactic fountain.  They point out, however, that most of the MS
detections are not consistent with the ionization model they use.

	More puzzling still is the detection by Weiner, et al. of
the HI associated with Sculptor, a dSph galaxy with a large cloud of HI
at a distance of 80 kpc.  This galaxy should
be a relatively good calibrator for the H$\alpha$ measurements because
the field-of-view covers a large fraction of the cloud, and because the
cloud has a well-determined distance.  Many of the H$\alpha$ detections
in Weiner et al. are factors of 2 -- 5 below the measured surface
brightness of Sculptor, suggesting that these clouds have distances of
100 - 200 kpc!

	Unfortunately, the interpretation of the H$\alpha$ measurements
seem to be fraught with uncertainty, and instead of giving the clean
result that had been hoped for, the measurements have raised numerous
questions in their own right.  So far, measurements have only been made
toward southern hemisphere HVCs, which according to the Local Group
hypothesis, should be closer than average, and some of them may be
contaminated by debris from the MS.  Another test would be to observe
the small HVCs with high negative velocities within a radian of M31,
which should have large distances from the MW and well-determined
distances from M31.  Where the projected distance from M31 is large,
the measured H$\alpha$ fluxes should be lower, on
average, than those observed in the southern hemisphere. 

\subsection {The Metallicity Test}

	An unambiguous prediction of the Local Group hypothesis is that
the HVCs should have substantially subsolar metallicities.  But how low is
appropriate?  If the HVCs are truly primordial, their metallicities should
be zero, but it has been difficult to find {\it any} intergalactic gas
with zero metallicity, especially for gas with column densities like that
of the HVCs: the Lyman limit systems.  Blitz et al. (1999) suggest values
$<$ 0.1 -- 0.3 solar based on various measurements of intergalactic gas.
HVCs originating in a Galactic fountain should have metallicities that are
{\it at least} solar because most come from regions interior to the Solar
distance and because they are accelerated by supernovae, stellar winds from
O stars, etc. from which they should get higher than solar metallicities.
In order to achieve high velocities relative to the LSR, these clouds must 
not become too well mixed with halo gas, and if this is the case, the
high metallicities will be maintained.

	Gibson (2001, this volume) summarizes most of the metallicity data
published to date (including some of his own unpublished data), and shows
that most of the HVCs do indeed typically have metallicities of 0.3
or less.  Many of them do not have measurements of the ionized
component, and may have metallicities lower than the tabulated value.
The highest metallicity cloud in Gibson's list is probably different
from all other catalogued HVCs: the original HI detection in the
is not confirmed in the Leiden-Dwingeloo HI survey.

	The most straightforward conclusion from the metallicity data 
is that the measurements are inconsistent with the Galactic fountain
model, seemingly fatal to it.  Even though the number of good
measurements are few, no {\it bona fide} HVC has either solar or
supersolar metallicity!  Combined with the results from the H$\alpha$
test, it seems that the Galactic fountain model can be ruled out as
being a significant contributor to the HVC phenomenon.  This does not
mean, of course, that the Galactic fountain doesn't exist.  Indeed,
Blitz et al. give other evidence for the existence of the Galactic
fountain, but it does not, apparently play an important role in the HVC
phenomenon.  The model, it seems, has to be fundamentally altered to
fit the data.

	Nevertheless, the metallicity data give values somewhat higher
than what might be naively expected from the Local Group hypothesis.  
Furthermore, all but one of the
HVCs that have been measured are in the southern hemisphere where there
may be some confusion with MS gas; the interpretation of the origin
of the metallicities is therefore not unambiguous one way or the other
as Gibson (this volume) points out.  As is true for the H$\alpha$ test,
it would be useful to have some of the small, high negative velocity
clouds near M31 measured, but finding suitable background sources is
difficult.

\subsection {The Extragalactic Analogue Test}

	In the absence of direct distance measurements, which probe
only the nearest clouds, 
the most direct test of the Local Group hypothesis
is to find analogues in other groups similar to the Local Group.  The number
density of HVC analogues ought to be related to environment, and it is
likely that in rich clusters, for example, the HVC analogues have been,
for the most part, accreted.  In these systems HVC analogues would still
be expected to occupy the outer reaches of a cluster, but perhaps
with rather low surface filling fraction.

	Estimates of the detectability of HVC analogues requires
knowledge of the size and column density (or mass) of the local HVCs.
Blitz et al. provided an estimate based on an assumed median distance
of 1 Mpc and no correction for beam convolution, which could not be
made from the Wakker \& van Woerden (1991) data.  The size and mass
estimates have some flexibility however; the median distance can be
as close as 500 kpc without producing difficulties for the dynamical
modeling (but requires a larger ratio of dark matter to baryons),
and beam smearing is clearly important for the small clouds that make
up the majority of the sample (e.g. Wakker \& van Woerden 1991;
Braun \& Burton 1999).  Assuming a
distance of 700 kpc and an estimate of the effect of beam smearing from
the data of Hartmann \& Burton (1997),  a typical HVC has a diameter
of about 14 kpc, a typical HI mass of about 5 $\times 10^6$ M\sun, and
a similar ensemble of HVCs in a distant group has a 
surface filling fraction on the sky of about 1\%.

	Even with the earlier, larger size and mass estimates, 
direct detection of HVC analogues by either emission line or absorption
line experiments would be difficult, as pointed out by Blitz et
al. because of the low surface filling fraction, and because of beam dilution,
except for relatively nearby groups.  Nevertheless, several 
sensitive HI surveys have been made, and the most sensitive of these, the
Arecibo HI Sky Survey (AHISS; Zwaan et al. 1997), failed to detect
any HVC analogues (Zwaan \& Briggs 2000). i
These authors argued that they should have
detected 70 HVCs around groups and about 250 HVCs around galaxies
in their survey based on the sizes and masses given by Blitz et al. (1999).
They concluded that if the HVCs are indeed related to the Milky Way,
they must have distances $<$ 200 kpc;  their conclusions cast considerable
doubt on the Local Group hypothesis. 

	The Zwaan \& Briggs result was, however, recently reevaluated by
Braun \& Burton (2001), who find several fundamental flaws.  First,
the assumed noise of the AHISS was found to be somewhat underestimated and the
sensitivity somewhat overestimated.  Second, most of the
groups and individual galaxies Zwaan \& Briggs considered are too far
away and the covering fraction of HVC analogues is 
too small for the AHISS non-detections to
place significant limits on the number of HVC analogues in these systems.
Third, Zwaan \& Briggs included any field galaxy that fell within 1 Mpc
of the AHISS survey strip in their analysis, but Braun \& Burton (2001)
argue reasonably that distance is too high by about an order of magnitude.
When Braun \& Burton (2001) make the necessary corrections, they find
only one group, the NGC 628 group, within a distance range that could put
significant constraints on the number of extragalactic HVC analogues.
Yet the number of clouds in that group that could be present and still
be consistent with the Zwaan \&
Briggs non-detections is comparable to to the number expected to be in the
Local Group from the
catalogue of Wakker \& van Woerden (1991)!  Braun \& Burton (2001) find
a similar result when they considered individual field galaxies.  Thus,
the non-detections in the AHISS survey do not place useful constraints
on HVC analogues in other systems.  The reason for the difference from
the Zwaan \& Briggs analysis is that Braun \& Burton (2001) show that
only in the nearest galaxies and groups are mass limits sufficiently
sensitive and in these groups the survey samples only a small fraction
of the relevant projected area of the sky.  Burton \& Braun (2001) go
further and examine all of the relevant HI surveys, and find that none of
those published to date put a limit the number of HVC analogues in other
systems inconsistent with a generalization of the Local Group hypothesis.

	To overcome some of the difficulties with using the AHISS,
Zwaan (2001) did a targeted survey of six galaxy groups with the
Arecibo telescope.  By doing Monte Carlo simulations of HVC analogues
within these groups, he concluded that  between 6 and 28 sources should
have been detected, depending on the assumptions, if the number of HVC
analogues in each group is 100.  Zwaan did detect several sources, but
argued that none are HVC analogues.  However, in none of the groups is the
fractional surface area covered by the observations larger than 0.005.
Based on a surface filling fraction of 0.01 in each group, the total
number of detections expected in Zwaan's data is two, just the
number of HI clouds he detects that are not associated with a galaxy.
These detections may be analogues of the nearby complexes A, C and M.
In any event, the number of extragalactic HVC analogues is still 
poorly constrained by the observations.

	Yet, even if none of the HI surveys to date can rule out that HVC
analogues are seen in other groups, shouldn't it be possible to find the
largest such systems in some other groups?  Why hasn't the upper end of
the HVC luminosity function been detected in other systems?  Part of
the answer is that the HVC luminosity function cannot be determined
from the data at hand because individual HVC distances are not known.
On the other hand, at least one  HVC analogue has now been identified in
the nearby Universe, with a mass of 1.7 $\times$ 10$^7$ M\sun, a diameter
of 15 kpc at an estimated distance of 3.2 Mpc, and no stars to a limiting
$\mu(B) \sim$ 27 mag arcsec$^{-2}$ (Kilborn et al 2000).  The distance,
based on a heliocentric velocity of 450 km s$^{-1}$ and the assumption
that it is in Hubble flow, is rather uncertain.  Thus, at least one HVC
analogue of mass and size within the range expected, and incidentally
requiring a large amount of dark matter to be stable,  has been identified,
but where are the others?  Blitz et al. (1999) catalogued a number of high
mass HVC analogues from the literature, but because of their proximity to
massive galaxies, it cannot be certain that these are not tidal features,
though in most cases, they do not have tidal morphologies.

	One solution is to look at the Sculptor group, the group 
nearest the Milky Way at a distance of about 1.5 Mpc.  Many of
the higher mass clouds should have been detected in the HIPASS Survey,
and in the southern extension of the Leiden-Dwingeloo HI survey (Arnal
et al. 2000).  However, the Sculptor group is situated behind the
Magellanic Stream and is confused in velocity at many positions with
it.  Nevertheless, the velocity dispersion of the group is much
larger than that of the HI in the MS and should be separable from it.
A tentative confirmation of an increased HI velocity dispersion in the
direction of Sculptor was published by Putman (2000), but the data have
not yet been analyzed in detail. The HIPASS survey has not been corrected
for stray Galactic radiation in the sidelobes and so is useful only at
velocities beyond those of the Galactic emission.  With its larger beam,
the Villa Elisa HI survey has a somewhat lower mass sensitivity than HIPASS, but
it can be corrected for sidelobe contamination and will be a good test.
Not finding HVC analogues in the Sculptor group, if these surveys are
as sensitive as are claimed, would likely prove fatal for the Local
Group hypothesis.

\subsection {The MgII Test}

	If HVC analogues populate groups of galaxies like the Local
Group, they should occasionally be seen in quasar absorption lines,
since these lines of sight are sensitive to much lower HI column
densities than 21-cm emission lines.  Recently, Charlton, Churchill \&
Rigby (2000) examined the statistics of of moderate redshift MgII and
Lyman limit absorbers in QSO absorption lines as a probe to see what
sort of contribution might come from HVC analogues.  

	Stripped to its essentials, the argument made by Charlton et
al. is as follows.  Strong MgII absorbers are found in 58 systems toward 51
quasars in a survey by Steidel, Dickinson \& Persson (1994).  However,
all but 3 of the 58 absorbers have identified galaxies with a
coincident redshift within 40h$^{-1}$ kpc of the quasar, so presumably all but
5\% of the strong MgII absorbers are from the galaxies, which are
generally normal and bright ($L \ge 0.1 L*$).  This leaves only a small
contribution possible from a population of HVC analogues.  However, the
covering fraction of HVC analogues, per galaxy group, is equal to 
$N_{HVC}/N_{gal} \times (R_{HVC}/R_{gal})^2$, where $N_{HVC}$
and $N_{gal}$ are the number of HVCs and galaxies within a particular
group, and where $R_{HVC}$ and $R_{gal}$ are mean radii of HVCs and
galaxies in the groups.  For the values of these quantities of 300, 4,
7.5 kpc and 40 kpc respectively, the HVC covering fraction is about 2.5
times that of galaxies.  Thus either the surface filling fraction of
MgII absorbing gas is $<< 1$, or the Local Group hypothesis
overpredicts the number of MgII absorbers.  Similar arguments are made
about weak MgII absorbers, and Lyman Limit systems.

	This simple, persuasive argument has at least one serious
weakness.  The criterion for positional coincidence in the Steidel et
al. (1994) survey is that the galaxy be within 40h$^{-1}$ kpc, or about 60
kpc, and the required velocity coincidence is several hundred km s$^{-1}$,
limited primarily by the precision of the galaxy redshifts.  The impact
parameter is, however,  much larger than a typical HI radius even for
a large galaxy.  Furthermore,  Dickinson \& Steidel (1996) find that
the absorbers are consistent with a {\it spherical} distribution in the
galaxies, rather than a disk-like distribution.  Thus it may be be that
a substantial fraction of the MgII absorbers are due not to the galaxy
itself, but to HVCs along the line of sight, either close to the galaxies
as an HVC is being accreted or in the intergroup gas of a parent galaxy.
If, for example, this is the case in about half the galaxies, then the
Charlton et al. constraint is considerably softened, and the statistics of
the MgII absorbers rather than providing a strong constraint against the
Local Group hypothesis, could instead provide important support for it.

\section {Other Distance Indicators}

	The availability of direct distance determinations remains
disappointingly sparse.  However, Braun \& Burton (2000) and Br\"uns,
Kerp, \& Pagels (2001) have suggested a new way to determine the distances
to the HVCs.  The basic idea is that in some cases, both the column
density and angular diameter of an HVC are well measured quantities;
aperture synthesis observations sometimes make it possible to identify
dense clumps in the HVCs.  If it is possible to determine the density
of the clumps independently, one can solve directly for the distance.
Braun \& Burton (2000) have used this technique to estimate the distances
to several clouds which are typically in the range of several hundred kpc,
strengthening the Local Group hypothesis.

	Both Braun \& Burton (2000) and Br\"uns et al. have found clumps with
linewidths so narrow that it is possible to get an upper limit to the
kinetic temperature of the clumps.  If the depth of the clump is
comparable to its dimensions on the sky, then its density and
therefore internal pressure depends only on its distance.  If the
external pressure is known, then the distance to the clump can be
found under the assumption of pressure equilibrium.  The difficult
part is getting a measure of the external pressure.  Braun \& Burton (2000)
use an extension of a model of Wolfire et al. (1995), but it is
unclear whether the model is applicable to the very low pressures of
the intergalactic medium.  The derived distances are therefore highly
model dependent.  Br\"uns et al. take a different tack and assume that
the clumps are virialized, an unjustified assumption for deriving the
distance.  If the assumptions in both of these models are correct,
then the distance determinations are probably sound.  However, because
the estimates are so dependent on the assumptions, they cannot be used
as primary distance discriminants for the HVCs.

\section {Summary Evaluation}   

	Of the four tests, two, the metallicity test and the H$\alpha$
test appear to rule out the Galactic fountain as contributing
significantly to the HVC phenomenon.  This leaves the Local Group
hypothesis, and either tidal or ram pressure stripping of gas from Local
Group galaxies as the main contenders for being the dominant origin of
the HVCs.  It could be that some of the HVCs are debris from the MS,
but clouds that are stripped from the Magellanic Clouds should lie
on a great circle with the MS, and most of the HVCs do not.  No other
galaxies in the Local Group have been identified as potential progenitors
for these clouds, because, if the HVCs are not self-gravitating, they
must be rather short lived.  Thus although some authors have suggested
that the HVCs might be tidal remnants (e.g.  Wakker et al. 1999), there
is neither kinematic nor dynamical evidence to support this idea.

	Searches for extragalactic HVC analogues have not
placed strong constraints on the number of HVCs with mean diameters of 15
kpc and mean masses of 5 $\times 10^6$ M\sun.  However, the larger mean
size and mass originally estimated by Blitz et al. (1999) is 
difficult to sustain in view of the results of the HI surveys of Zwaan \&
Briggs (2000) and Zwaan (2001), if the number of HVCs is as large as that
implied in the catalogue of Wakker \& van Woerden (1991).  However, it is
also possible, that many objects identified as tidal features in groups
of galaxies and even near field galaxies, are actually extragalactic HVC
analogues and do not have their origins in galaxies.  These may plausibly
be  extragalactic analogues of complexes A, C and H, the HVCs closest to
the Milky Way and probably in the process of being accreted.  After all,
if complex C were viewed from, say, M81 with telescopes comparable to
what we have been using, its long stringy morphology in close proximity
to the MW would be very suggestive of a tidal feature.

	The dynamical evidence remains the best evidence for the Local
Group hypothesis, and although the various tests are not in obvious
contradiction to it, neither do they provide strong confirmation.
Rather, the H$\alpha$ observations remain puzzling, and the
MgII absorbers  provide an important constraint only if the gas associated
with these systems is much more extended and is distributed spherically,
both of which are very different from the gas seen in spiral galaxies at
zero redshift.  Although the metallicities measured so far fall within
the range predicted by Blitz et al. (1999), they do remain uncomfortably
high, higher then typical metallicities measured in LG dwarf spheroidal
galaxies, for example.  Nevertheless, if the Local Group hypothesis turns
out to be incorrect, it will be challenging for an alternate hypothesis
to produce a good simple explanation for both the kinematic data and the
spatial distribution of the HVCs, which has been where other ideas have
always been the weakest.

% For examples on including figures, see the file vla2000_sample.ps
% at http://www.nrao.edu/vla2000/proceedings/. 
% For examples of figures, equations or tables, please see the file
% vla2000_man.ps at the same site. Also available as
% newpaspman.ps at http://www.aspsky.org/pubs/authors.html

% comment this out if you want to include acknowledgements

%\acknowledgements

\end{document}